\documentclass{article}%
\usepackage{amsmath}
\usepackage{amsfonts}
\usepackage{amssymb}
\usepackage{graphicx}%
\providecommand{\U}[1]{\protect\rule{.1in}{.1in}}

\begin{document}

\title{Subdynamics of a many-particle classical system driven from an\\equilibrium state by an external force}
\author{Victor F Los
\and Institute for Magnetism, Nat. Acad. Sci.and Min. Edu. Sci.
\and of Ukraine, 36-b Vernadsky Blvd., 03142 Kiev, Ukraine
\and victorflos43@gmail.com
\and tel. 067-500-7817 }
\maketitle

\begin{abstract}
It is shown, that by means of a special projection operator, the

Liouville equation for an $N$-particle distribution function of classical

particles, driven from an equilibrium state by an external field,

can be exactly converted into a closed linear homogeneous

Generalized Master Equations (GMEs) for an $s$-particle ($s<N$)

distribution function. The obtained linear time-convolution and

time-convolutionless GMEs define a subdynamics in the $s$-particle

phase space and contains no inhomogeneous initial correlations terms

as compared to the conventional GMEs. No approximation like

"molecular chaos" or Bogoliubov's principle of weakening of initial

correlations is needed. The initial correlations are "hidden" in the

projection operator and thus they are accounted for in the obtained

equations. For the weak interparticle interaction and weak external

field, these equations are rewritten in the second order of the

perturbation theory. Essentially, that they contain the contribution of

initial correlations in the kernel governing the evolution of an $s$-particle

distribution function. In particular, the evolution equation for a

one-particle distribution function is obtained and its connection to the

nonlinear Landau and the Fokker-Planck equations is discussed. The

obtained results are related to the general issues of statistical physics

and to the physical applications (plasma physics).

\end{abstract}

\section{Introduction}

One of the basic tasks of the nonequilibrium statistical physics remains the
deriving of the appropriate evolution equations for the measurable values
(statistical expectations) characterizing a nonequilibrium state of a
many-particle system. It is expected that these equations should generally be
evolution equations converting into kinetic (or other irreversible) equations
after taking some scaling limits. The principal question is: How to derive
such equations rigorously from underlying microscopic reversible classical or
quantum dynamical equations? Several approaches are usually used to address
this problem commonly starting with the Liouville equation for a distribution
function (classical mechanics) or the von-Neumann equation for a statistical
operator (quantum mechanics) for the $N$-particle ($N\gg1$) system under
consideration. The problem is that it is impossible to resolve these equations
due to a huge number of variables on which a distribution function
(statistical operator) depends. Fortunately, we do not need such an excessive
information, and a distribution function (statistical operator) for a much
smaller $s<<N$ number of particles is necessary for calculation of the
measurable values of interest. But due to the interparticle interaction, the
strict derivation of the completely closed equations for these reduced
distribution functions (marginals) is still a challenging problem. The
approximations allowing to obtain closed evolution equations for marginals,
such as the Bogoliubov principle of weakening of initial correlations
\cite{Bogoliubov (1946)} or RPA, actually imply selecting the specific
uncorrelated (factorized) distribution function (statistical operator) as an
initial state of the system, and then the problem of the "propagation of
chaos" with time should be resolved. Strict justification of the propagation
of "molecular chaos" turned out to be not an easy task. It was supposed to be
true in the so-called "Boltzmann-Grad limit" \cite{Grad1, Grad2}. It is also
worth noting that the nonlinearity of the kinetic equations, derived from the
linear Liouville (or von-Neumann) equation, comes from the nonlinear
(factorized) initial state (\cite{Kac}), which is not a realistic one
\cite{van Kampen (2004)}.

In the approach leading to the so called generalized master equations (GMEs),
an $N$-particle distribution function $F_{N}(x_{1},\ldots,x_{N};t)$ (we will
focus on the classical physics case when $x_{i}=(\mathbf{x}_{i},\mathbf{p}%
_{i})$ is the coordinate of the $i-$th particle in the phase space) is divided
into the essential (relevant) $\rho_{r}(x_{1},\ldots,x_{N};t)$ and inessential
(irrelevant) $\rho_{i}(x_{1},\ldots,x_{N};t)$ parts by using some projection
operators $P$ and $Q=1-P$. \ Applying the projectors $P$ and $Q$ to the
Liouville equation, one can obtain the time-convolution (non-Markovian) GME
(TC-GME) \cite{Nakajima (1958)}, \cite{Zwanzig (1960)}, \cite{Prigogine
(1962)} and time-convolutionless (time-local) GME (TCL-GME) \cite{Shibata1
(1977)}, \cite{Shibata2 (1980)}, which are the exact equations for the
relevant part of a distribution function usually selected as a part of a
distribution function with no correlations between the selected (by $P$)
subsystem and remaining part of the system. However, these equations are the
inhomogeneous ones, i.e., include a source (irrelevant part) containing all
many-particle correlations at the initial instant $t_{0}$. In general, these
initial correlations pose a problem to deal with, and Bogoliubov's principle
of weakening of initial correlations or simply the factorized initial
conditions (RPA) are commonly used to exclude initial many-particle
correlations (a source). These approximations result in an approximate closed
linear homogeneous GMEs for the relevant part of the distribution function.

The GMEs approach is standard for the important case of a subsystem $S$ of the
total system interacting with the remaining part $B$ of a full system in an
equilibrium state (a thermal bath). In this case, the factorizing initial
condition for the distribution function $F_{N}(t)$ of the whole $N$-particle
system
\begin{equation}
F_{N}(t_{0})=F_{S}(t_{0})F_{B}^{eq} \label{2}%
\end{equation}
is conveniently used ($F_{S}(t)=%
{\displaystyle\int}
dx^{B}F_{N}(t)$ is a subsystem distribution function defined as the integrated
over the thermal bath states the full $N$-particle distribution function,
$F_{B}^{eq}$ is the bath equilibrium distribution function), and the standard
projection operator for this case is defined as%
\begin{equation}
P(...)=F_{B}^{eq}\int dx^{B}(...). \label{3}%
\end{equation}

On the other hand, the Prigogine school was developing an approach (see, e.g.,
\cite{Balescu}) based on the introducing of the projection operator selecting
the kinetic part of a distribution function (statistical operator) of a
many-particle system. The projector is defined in such a way that the
evolution of this kinetic component is described by the closed homogeneous
kinetic equation (with no irrelevant initial moment source term) and this
process is termed as subdynamics. Such an approach is rather complicated,
depends on several unproved conditions, and no specific projection operator of
this type was identified. Moreover, the factorized initial state of the whole
system was again used in order to obtain the nonlinear kinetic equation
\cite{Balescu}.

In this paper, we apply the GME approach to considering an evolution in time
of a classical many-particle system driven by an external force from an
initial equilibrium state of the whole system. We show, that there is a
specific projection operator selecting the relevant part of the $N$-particle
distribution function which is governed by the exact homogeneous closed linear
GMEs. The initial correlations are comprised into the projection operator.
These equations are equivalent to the closed evolution equations for the
reduced $s$-particle ($s<N$) distribution function. For a small interparticle
interaction, the projection operator can be expanded in the interaction
series, and thus the corrections, given by initial correlations, can be found.
Therefore, we provide the example of the existence of the operator exactly
projecting the $N$-particle dynamics on the $s$-particle one and suggest the
method of accounting for initial correlation. We apply this approach to the
weakly interacting system of classical particles. In particular, the evolution
equation for the one-particle distribution function accounting for initial
correlations is obtained in the second order approximation in the
interparticle interaction and for a weak external field. The connection of
this equation to the nonlinear Landau equation and to the Fokker-Planck
equation for the weakly interacting gas of classical particles in the
space-homogeneous case is discussed. The obtained results seems important for
the general problems of the statistical physics as well as for the physical
applications, e.g., for the plasma physics.

\section{General formalism}

Let us consider the Liouville equation for a distribution function
$F_{N}(t,t_{0})$ of $N$ interacting classical particles%
\begin{align}
\frac{\partial}{\partial t}F_{N}(t,t_{0})  &  =-L(t)F_{N}(t,t_{0}),\nonumber\\
\int...\int dx^{N}F_{N}(t,t_{0})  &  =1,dx^{N}=dx_{1}...dx_{N}, \label{4}%
\end{align}
where $F_{N}(t,t_{0})=F_{N}(x_{1},...,x_{N};t,t_{0})$ is a function of $N$
variables $x_{i}=(\mathbf{r}_{i},\mathbf{p}_{i})$ ($i=1,...,N$) representing
the coordinates and momentum of the particles, $L(t)$ is the Liouville
operator acting on $F_{N}(t,t_{0})$ as%

\begin{equation}
L(t)F_{N}(t,t_{0})=\{H(t),F_{N}(t,t_{0})\}_{P}=\sum_{i=1}^{N}\{\frac{\partial
F_{N}(t,t_{0})}{\partial\mathbf{r}_{i}}\frac{\partial H(t)}{\partial
\mathbf{p}_{i}}-\frac{\partial F_{N}(t,t_{0})}{\partial\mathbf{p}_{i}}%
\frac{H(t)}{\partial\mathbf{r}_{i}}\}. \label{5}%
\end{equation}
Here, $\{H,F_{N}(t,t_{0})\}_{P}$ is the Poisson bracket and $H(t)$ is the
Hamilton function for the system under consideration generally dependent on time.

The formal solution to Eq. (\ref{4}) is
\begin{equation}
F_{N}(t,t_{0})=U(t,t_{0})F_{N}(t_{0},t_{0}), \label{6}%
\end{equation}
where the evolution operator $U(t,t_{0})$ is defined as
\begin{align}
U(t,t_{0})  &  =\exp[-\int\limits_{t_{0}}^{t}d\xi L(\xi)],\nonumber\\
U(t_{0},t_{0})  &  =1. \label{7}%
\end{align}

It is practically impossible to solve Eq. (\ref{4}) for a many-particle
system. Fortunately, however, in order to calculate the measurable values
(statistical expectations) of interest, one usually only needs to know the
reduced distribution functions (marginals) $F_{s}(t,t_{0})=F_{s}%
(x_{1},...,x_{s};t,t_{0})$ dependent on much smaller number of variables
$s<<N$. In general, the $s$-particle ($s\leq N$) distribution function is
defined as \cite{Bogoliubov (1946)}
\begin{equation}
F_{s}(t,t_{0})=V^{s}\int\cdots\int dx^{N-s}F_{N}(t.t_{0}),dx^{N-s}%
=dx_{s+1}...dx_{N}, \label{7a}%
\end{equation}
where $V$ is the volume of a system. From (\ref{4}), we have the normalization
condition for the reduced distribution functions $F_{s}$
\begin{equation}
\int\cdots\int dx^{s}F_{s}(t.t_{0})=V^{s},dx^{s}=dx_{1}...dx_{s}. \label{7b}%
\end{equation}
Thus, an average value of a function of the dynamic variables of the group of
$s$ particle is defined by the reduced distribution function $F_{s}$ as%
\begin{equation}
<A_{s}>_{t,t_{0}}=\int...\int dx^{N}A_{s}F_{N}(t,t_{0})=\int...\int
dx^{s}A_{s}\frac{1}{V^{s}}F_{s}(t,t_{0}). \label{7c}%
\end{equation}

In order to obtain equations for the reduced distribution functions, it is
convenient to employ the standard projection operator technique \cite{Nakajima
(1958)}, \cite{Zwanzig (1960)}, \cite{Prigogine (1962)} and to break
$F_{N}(t,t_{0})$ into the relevant $\rho_{r}^{s}(t,t_{0})$ and irrelevant
$\rho_{i}(t,t_{0})$ parts%
\begin{align}
F_{N}(t,t_{0})  &  =\rho_{r}^{s}(t,t_{0})+\rho_{i}(t,t_{0}),\nonumber\\
\rho_{r}^{s}(t,t_{0})  &  =PF_{N}(t,t_{0}),\rho_{i}(t,t_{0})=QF_{N}%
(t,t_{0})=F_{N}(t,t_{0})-\rho_{r}^{s}(t,t_{0}) \label{8}%
\end{align}
with the help of some projection operators $P$ and $Q=1-P$ ($P^{2}=P$,
$Q^{2}=Q$, $P+Q=1$, $PQ=0$). We note, that the relevant and irrelevant parts
depend on coordinates and momenta of all $N$ particles in contrast to the
reduced distribution functions (like $F_{s}(t,t_{0})$). The relevant part
$\rho_{r}^{s}(t,t_{0})$ is conveniently defined in such a way that it is
related to the reduced distribution function of interest $F_{s}(t,t_{0})$,
i.e., the projection operator selects a relevant ("vacuum") part of
$F_{N}(t,t_{0})$, which contains no correlation between a subsystem (a group
of $s$ particles) and an environment (remaining $N-s$ particles), and actually
describes the evolution of $F_{s}(t,t_{0})$. Then, the irrelevant part of the
distribution function $\rho_{i}(t,t_{0})$ contains all correlations between a
subsystem and an environment.

Applying the projection operators $P$ and $Q$ to Eq. (\ref{4}), it is easy to
obtain the equations for the relevant and irrelevant parts of $F_{N}(t)$
\begin{align}
\frac{\partial}{\partial t}\rho_{r}^{s}(t)  &  =-PL(t)[\rho_{r}^{s}%
(t)+\rho_{i}(t)],\nonumber\\
\frac{\partial}{\partial t}\rho_{i}(t)  &  =-QL(t)[\rho_{r}^{s}(t)+\rho
_{i}(t)] \label{9}%
\end{align}
(from now on we put $t_{0}=0$). A formal solution to the second Eq. (\ref{9})
has the form
\begin{align}
\rho_{i}(t)  &  =-%
{\displaystyle\int\limits_{0}^{t}}
U_{Q}(t,\tau)QL(\tau)\rho_{r}^{s}(\tau)d\tau+U_{Q}(t,0)\rho_{i}(0),\nonumber\\
U_{Q}(t,\tau)  &  =\exp[-%
{\displaystyle\int\limits_{\tau}^{t}}
d\xi QL(\xi)Q],\nonumber\\
\rho_{i}(0)  &  =F_{N}(0)-PF_{N}(0),F_{N}(0)=F_{N}(0,0). \label{10}%
\end{align}
Inserting this solution into the first Eq. (\ref{9}), we obtain the
conventional exact time-convolution generalized master equation (TC-GME) known
as the Nakajima-Zwanzig equation for the relevant part of the distribution
function
\begin{align}
\frac{\partial}{\partial t}\rho_{r}^{s}(t)  &  =-PL(t)\rho_{r}^{s}%
(t)+\int\limits_{0}^{t}PL(t)U_{Q}(t,\tau)QL(\tau)\rho_{r}^{s}(\tau
)d\tau\nonumber\\
&  -PL(t)U_{Q}(t,0)\rho_{i}(0). \label{11}%
\end{align}

This equation is quite general and valid for any initial distribution function
$F_{N}(0)$. Serving as a basis for many applications, Eq. (\ref{11}),
nevertheless, contains the undesirable and in general non-negligible
inhomogeneous term (the last term in the right hand side of (\ref{11})), which
depends via $\rho_{i\text{ }}(0)$ on the same large number of variables as the
distribution function at the initial instant $F_{N}(0)$ and includes all
initial correlations. Therefore, Eq. (\ref{11}) does not provide for a
complete reduced description of a multiparticle system in terms of the
relevant (reduced) distribution function. Applying Bogoliubov's principle of
weakening of initial correlations (allowing to eliminate the influence of
$\rho_{i}(0)$ on the large enough time scale $t\gg t_{cor}$) or using a
factorized initial condition (see (\ref{2})),when $\rho_{i\text{ }}%
(0)=QF_{N}(0)=0$ (i.e., e.g., $F_{N}(0)=\rho_{r}^{s}(0)$), one can achieve the
above-mentioned goal and obtain the homogeneous GME for $\rho_{r}^{s}(t)$,
i.e. Eq. (\ref{11}) with no initial condition term. However, obtained in such
a way homogeneous GME is either approximate and valid only on a large enough
time scale (when all initial correlations vanish) or applicable only for a
rather artificial (actually unreal, as pointed in \cite{van Kampen (2004)})
initial conditions (no correlations at an initial instant of time). In
addition, Eq. (\ref{11}) poses the problem to deal with due to its
time-nonlocality. However, it is possible to obtain the time-local equation
for the relevant part of the subsystem distribution function \cite{Breuer,
Shibata1 (1977), Shibata2 (1980)} (see also below).

Then, the interesting question arises: Is it possible to obtain exact and
completely closed (homogeneous) GME, i.e. the equation with no inhomogenous
initial correlations \ term? It means that the initial correlations would be
contained in the kernel governing the evolution of the relevant part of the
distribution function. As we already mentioned in the Introduction, there are
the ways to include initial correlations into consideration on an equal
footing with collisions (see \cite{Los JPA, Los JSP, Los TMP, Los Evolution
Equations}.

The initial equilibrium state for the whole system provides the new
opportunities for obtaining the homogeneous GMEs. It was demonstrated in the
recent works \cite{Los JSP 2017, Los Physica A 2018}), where the quantum case
of the subsystem (like an electron) interacting with a heat bath (like the
equilibrium boson field) was considered and the projection operator $P$ is of
a standard form (the quantum version of (\ref{3})) was employed.

\section{Completely closed (homogeneous) equations for a reduced distribution
function}

One can consider the following problem: Is it possible to introduce such a
projection operator which allows for converting the Liouville equation
(\ref{4}) into a completely closed (homogeneous) GME? If so, then this
projection operator should comprise in some way the initial correlations. This
idea is reminiscent of an approach, which had been developing by Prigogine
with coworkers (see, e.g. \cite{Balescu}). They assert that it is possible to
formally introduce some projection operator $\Pi$ allowing for the exact
dividing of the $N$-particle distribution function into the kinetic $f_{k}(t)$
and nonkinetic parts (both include "vacuum" and correlated terms), and the
"vacuum" part of the kinetic part $Vf_{k}(t)$ (the projection operator $V$
selects the "vacuum" part of $f_{k}(t)$ with no correlations) satisfies the
completely closed (homogeneous) kinetic (irreversible) equation like Eq.
(\ref{11}) but with no source term (the third term on the r.h.s. of
(\ref{11})) and the infinite ($\infty$) upper limit of the integration over
$\tau$. Such an evolution of $Vf_{k}(t)$ was termed subdynamics. However, this
approach is very formal, includes some not generally proven conditions, and
contains no explicit form of the projection operators $\Pi$ and $V$ for an
arbitrary $N$-particle system.

Now we will show for the $N$-particle classical system driven from an initial
equilibrium state $\rho_{eq}$ by an external force that there is an explicit
projection operator enabling one to obtain homogeneous closed time-convolution
and time-convolutionless (time local) generalized master equations for a
relevant part of $N$-particle distribution function.

Thus, we now suppose that up to the moment of time $t=0$ the system is in an
equilibrium state with the Gibbs distribution function%

\begin{equation}
F_{N}(t\leq0)=\rho_{eq}=Z^{-1}\exp(-\beta H),\beta=1/k_{B}T,Z=\int...\int
dx^{N}\exp(-\beta H), \label{12}%
\end{equation}
but just after $t=0$ (at $t>0$) an external force is applied to a system. For
what follows, it is convenient to present the system's Hamilton function as%
\begin{align}
H(t)  &  =H_{s}+H_{\Sigma}+\widetilde{H}_{s\Sigma}+H^{F}(t),\nonumber\\
H^{F}(t)  &  =0,t\leq0. \label{13}%
\end{align}
\ Here, we selected the group of $s$ particles (described by $H_{s})$, which
interacts (through $\widetilde{H}_{s\Sigma}$) with an environment $\Sigma$ of
other $N-s$ particles (described by $H_{\Sigma}$), and $H^{F}$ defines the
influence of an external field.

Then, we the Liouville equation (\ref{4}) can \ be written as
\begin{align}
\frac{\partial}{\partial t}F_{N}(t)  &  =-L(t)F_{N}(t),t>0,\nonumber\\
\frac{\partial}{\partial t}F_{N}(t)  &  =0,t\leq0, \label{13a}%
\end{align}
and the Liouville operator $L(t)=L_{s}+L_{\Sigma}+\widetilde{L}_{s\Sigma
}+L^{F}(t)$. The formal solution to Eq. (\ref{13a}) is
\begin{align}
F_{N}(t)  &  =U(t)\rho_{eq},t\geq0,\nonumber\\
F_{N}(0)  &  =\rho_{eq}, \label{13b}%
\end{align}
where the evolution operator $U(t)$ is defined by (\ref{7}) (at $t_{0}=0$).

Conventionally, a natural choice for a projection operator in Eq. (\ref{11})
is the operator of the type given by (\ref{3}). For a system of classical
particles under consideration, such projection operator can be defined as%
\begin{align}
P(...)  &  =\rho_{\Sigma}\int...\int dx^{N-s}(...),Q(...)=1-P(...),\nonumber\\
\rho_{\Sigma}  &  =\frac{1}{Z_{\Sigma}}\exp(-\beta H_{\Sigma}),Z_{\Sigma}%
=\int...\int dx^{N-s}\exp(-\beta H_{\Sigma}),dx^{N-s}=dx_{s+1}...dx_{N},
\label{14}%
\end{align}
and the relevant and irrelevant parts of $F_{N}(t)$ are%
\begin{align}
\rho_{r}^{s}(t)  &  =\rho_{\Sigma}\frac{1}{V^{s}}F_{S}(t),\nonumber\\
\rho_{i}(t)  &  =F_{N}(t)-\rho_{\Sigma}\frac{1}{V^{s}}F_{S}(t). \label{15}%
\end{align}
We see from Eqs. (\ref{11}) (\ref{12}), that for such a choice of the
projector, $\rho_{i}(0)=\rho_{eq}-\rho_{\Sigma}\frac{1}{V^{s}}F_{S}(0)\neq0$.
It is also worth noting, that formal introducing of the distribution function
$\rho_{\Sigma}$ in (\ref{14}) does not generally mean that the environment of
$N-s$ particles is in the equilibrium state.

Let us now introduce the following projection operators $P_{s}$ and $Q_{s}$
\begin{align}
P_{s}(...)  &  =\rho_{\Sigma}^{s}\int...\int dx^{N-s}(...),Q_{s}%
(...)=1-P_{s}(...),\nonumber\\
\rho_{\Sigma}^{s}  &  =\frac{1}{Z_{\Sigma}^{s}}\exp[-\beta(H_{\Sigma
}+\widetilde{H}_{s\Sigma})],\nonumber\\
Z_{\Sigma}^{s}  &  =\int...\int dx^{N-s}\exp[-\beta(H_{\Sigma}+\widetilde
{H}_{s\Sigma})]. \label{16}%
\end{align}
It is not difficult to see that $P_{s}^{2}=P_{s}$, $Q_{s}^{2}=Q_{s}$,
$P_{s}Q_{s}=0$. Then, we can divide $F_{N}(t)$ into the relevant $f_{r}%
^{s}(t)$ and irrelevant $f_{i}(t)$\ components as (compare with (\ref{15}))\ %

\begin{align}
F_{N}(t)  &  =f_{r}^{s}(t)+f_{i}(t)\nonumber\\
f_{r}^{s}(t)  &  =P_{s}F_{N}(t)=\rho_{\Sigma}^{s}\frac{1}{V^{s}}%
F_{S}(t),\nonumber\\
f_{i}(t)  &  =Q_{s}F_{N}(t)=F_{N}(t)-\rho_{\Sigma}^{s}\frac{1}{V^{s}}F_{S}(t)
\label{17}%
\end{align}
Note, that in the case of the projector (\ref{16}), we use the notations
$f_{r}^{s}(t)$ and $f_{i}(t)$ for the relevant and irrelevant components,
correspondingly, while for the conventional projector $P$ (\ref{14}) we leave
the notations $\rho_{r}^{s}(t)$, $\rho_{i}(t)$ and Eq. (\ref{11}). It is not
difficult to see that the dynamics of the average value (\ref{7c}) is
completely defined by the relevant part $f_{r}^{s}(t)$ of $F_{N}(t)$, i.e.,%
\begin{equation}
<A_{S}>_{t}=\int...\int dx^{s}A_{s}\frac{1}{V^{s}}F_{S}(t)=\int...\int
dx^{N}A_{s}f_{r}^{s}(t). \label{18}%
\end{equation}

\bigskip The projection operator $P_{s}$ (\ref{16}) has an interesting
property, namely,%
\begin{equation}
P_{s}\rho_{eq}=\rho_{eq},Q_{s}F_{N}(0)=0. \label{19}%
\end{equation}

\subsection{Time-convolution homogeneous GME}

Thus, \bigskip by applying the introduced projection operators $P_{s}$ and
$Q_{s}$ to the Liouville equation (\ref{13a}), we arrive at the following
homogeneous time-convolution GME (compare with (\ref{11}))%
\begin{align}
\frac{\partial}{\partial t}f_{r}^{s}(t)  &  =-P_{s}L(t)f_{r}^{s}%
(t)+\int\limits_{0}^{t}P_{s}L(t)U_{Q_{s}}(t,\tau)Q_{s}L(\tau)f_{r}^{s}%
(\tau)d\tau,\nonumber\\
U_{Q_{s}}(t,\tau)  &  =\exp[-%
{\displaystyle\int\limits_{\tau}^{t}}
d\xi Q_{s}L(\xi)]. \label{20}%
\end{align}
\bigskip Equation (\ref{20}) shows, that in the considered case, the dynamics
of the $N$-particle distribution function can be exactly projected on the
dynamics within its relevant part subspace. It follows, that the evolution of
the selected complex of $s$ particles can be described by the \textbf{linear}
equation in the subspace of the corresponding coordinates $x_{i}%
=(\mathbf{r}_{i},\mathbf{p}_{i})$ ($i=1,...,s$) if we rewrite Eq. (\ref{20})
as the equation for an $s$-particle distribution function
\begin{equation}
\frac{\partial}{\partial t}F_{S}(t)=-\left[  \int...%
{\displaystyle\int}
dx^{N-s}L(t)\rho_{\Sigma}^{s}\right]  F_{S}(t)+\left[  \int...%
{\displaystyle\int}
dx^{N-s}L(t)\int\limits_{0}^{t}d\tau U_{Q_{s}}(t,\tau)Q_{s}L(\tau)\rho
_{\Sigma}^{s}\right]  F_{S}(\tau). \label{21}%
\end{equation}

\subsection{Time-convolutionless homogeneous GME}

Generally, the evolution equation (\ref{20}) poses some problem to deal with
due to its time-nonlocality. It is possible, however, to obtain the exact
homogeneous time-local equation for the relevant part of the distribution
function. The idea is to take advantage of the evolution of the distribution
function, defined by (\ref{13b}), which leads to the relation
\begin{align}
F_{N}(\tau)  &  =U^{-1}(t,\tau)F_{N}(t),\nonumber\\
U^{-1}(t,\tau)  &  =\exp[%
{\displaystyle\int\limits_{\tau}^{t}}
d\xi L(\xi)], \label{22}%
\end{align}
Using (\ref{22}) and the conventional projection operator (\ref{14}), the well
known time-convolutionless equation for a relevant distribution function
$\rho_{r}^{s}(t)$, which contains the undesirable inhomogeneous term $\rho
_{i}(0)$ comprising the initial correlations, can be obtained.(see
\cite{Breuer, Shibata1 (1977), Shibata2 (1980)}). We will show now, that the
use of the projector (\ref{16}) instead of (\ref{14}) leads to the completely
closed homogeneous time-convolutionless GME for the relevant part of the
distribution function. We will briefly conduct the derivation which is a
rather standard one. First, we apply the projector (\ref{16}) to (\ref{22})
and obtain
\begin{equation}
f_{r}^{s}(\tau)=P_{s}U^{-1}(t,\tau)[f_{r}^{s}(t)+f_{i}(t)]. \label{23}%
\end{equation}
We also have the equation for the irrelevant part $f_{i}(t)$ (see (\ref{10})
and (\ref{19}))%
\begin{equation}
f_{i}(t)=-%
{\displaystyle\int\limits_{0}^{t}}
U_{Q_{s}}(t,\tau)Q_{s}L(\tau)f_{r}^{s}(\tau)d\tau, \label{24}%
\end{equation}
where $U_{Q_{s}}(t,\tau)$ is given by (\ref{20}). From two equations
(\ref{23}) and (\ref{24}) one finds that
\begin{align}
f_{i}(t)  &  =[1-\alpha(t)]^{-1}\alpha(t)f_{r}^{s}(t),\nonumber\\
\alpha(t)  &  =-%
{\displaystyle\int\limits_{0}^{t}}
U_{Q_{s}}(t,\tau)Q_{s}L(\tau)P_{s}U^{-1}(t,\tau)d\tau. \label{25}%
\end{align}
Substituting $f_{i}(t)$ (\ref{25}) into the \ projected by $P_{s}$ equation
(\ref{13a})%
\begin{equation}
\frac{\partial}{\partial t}f_{r}^{s}(t)=-P_{s}L(t)[f_{r}^{s}(t)+f_{i}(t)],
\label{26}%
\end{equation}
we finally obtain%
\begin{equation}
\frac{\partial}{\partial t}f_{r}^{s}(t)=-P_{s}L(t)[1-\alpha(t)]^{-1}f_{r}%
^{s}(t). \label{27}%
\end{equation}
\qquad If it is possible to expand the operator $[1-\alpha(t)]^{-1}$ into the
series in $\alpha(t)$, then the first two terms of this expansion results in
the following time-local equation (compare with (\ref{20}))%
\begin{equation}
\frac{\partial}{\partial t}f_{r}^{s}(t)=-P_{s}L(t)f_{r}^{s}(t)+P_{s}L(t)%
{\displaystyle\int\limits_{0}^{t}}
d\tau U_{Q_{s}}(t,\tau)Q_{s}L(\tau)P_{s}U^{-1}(t,\tau)f_{r}^{s}(t). \label{28}%
\end{equation}

Equations (\ref{20}) and (\ref{27}) present the main result of this section.
They show that the projector (\ref{16}) allows for selecting the relevant part
$f_{r}^{s}(t)$ of the multiparticle distribution function $F_{N}(t)$ which
satisfies the completely closed linear time-convolution and
time-convolutionless equations when a system is driven from an initial
equilibrium state (\ref{12}) by an external force. They, in fact, describe the
evolution of the $s$-particles marginals on the arbitrary timescale. Thus, one
remains in the scope of the linear evolution given by the Liouville equation
(\ref{4}) but should pay for this simplification by accounting for initial
correlations, which are conveniently ignored. It is also worth noting that the
developed formalism only works in the framework of classical physics (when the
terms of the Hamilton function (\ref{13}) commutes with each other). For
quantum physics a different approach is needed (see \cite{Los JSP 2017, Los
Physica A 2018}).

\section{Simplified homogeneous GMEs}

Let us specify the Hamilton function $H(t)$ (\ref{13}) for the case of the
identical particles with the two-body interparticle interaction $V_{ij}$ as%
\begin{align}
H(t)  &  =H_{S}+H_{\Sigma}+\widetilde{H}_{s\Sigma}+H^{F}(t),\nonumber\\
H_{s}  &  =\sum\limits_{i=1}^{s}\frac{\mathbf{p}_{i}^{2}}{2m}+\sum
\limits_{1\leq i<j\leq s}V_{ij}(\left\vert \mathbf{r}_{i}-\mathbf{r}%
_{j}\right\vert )+<H_{s\Sigma}>_{\Sigma},\nonumber\\
H_{\Sigma}  &  =%
{\displaystyle\sum\limits_{i=s+1}^{N}}
\frac{\mathbf{p}_{i}^{2}}{2m}+\sum\limits_{s+1\leq i<j\leq N}V_{ij}(\left\vert
\mathbf{r}_{i}-\mathbf{r}_{j}\right\vert ),\nonumber\\
\widetilde{H}_{s\Sigma}  &  =H_{s\Sigma}-<H_{s\Sigma}>_{\Sigma},H_{s\Sigma
}=\sum\limits_{i=1}^{s}\sum\limits_{j=s+1}^{N}V_{ij}(\left\vert \mathbf{r}%
_{i}-\mathbf{r}_{j}\right\vert ),\nonumber\\
H^{F}(t)  &  =\sum\limits_{i=i}^{N}V_{i}(\mathbf{r}_{i},t),t>t_{0}%
,H^{F}(t)=0,t\leq t_{0}.\label{29}\\
&  .\nonumber
\end{align}
Here, for simplicity, we take an external force $H^{F}(t)$ dependent only on
$\mathbf{r}_{i}$, and, for convenience, introduce the energy of the mean field
$<H_{s\Sigma}>_{\Sigma}$acting on the $s$-complex by the "equilibrium"
environment%
\begin{equation}
<H_{s\Sigma}>_{\Sigma}=\int...%
{\displaystyle\int}
dx^{N-s}\rho_{\Sigma}H_{s\Sigma}, \label{30}%
\end{equation}
where $\rho_{\Sigma}$ is given by (\ref{14}). Note, that $<H_{s\Sigma
}>_{\Sigma}$depends only on the coordinates of $s$ selected particles
$\mathbf{r}_{i}$ ($i=1,...,s$). For a space-homogeneous case, this mean field
does not depend on $\mathbf{r}_{i}$ ($i=1,...,s$).

The corresponding to (\ref{29}) Liouville operator $L(t)$ is%
\begin{align}
L(t)  &  =L_{s}+L_{\Sigma}+\widetilde{L}_{s\Sigma}+L^{F}(t),\nonumber\\
L_{s}  &  =%
{\displaystyle\sum\limits_{i=1}^{s}}
[\mathbf{v}_{i}\mathbf{\nabla}_{i}-(\mathbf{\nabla}_{i}<H_{s\Sigma}>_{\Sigma
})\frac{\partial}{\partial\mathbf{p}_{i}}]-\sum\limits_{1\leq i<j\leq
s}(\mathbf{\nabla}_{i}V_{ij})\cdot(\frac{\partial}{\partial\mathbf{p}_{i}%
}-\frac{\partial}{\partial\mathbf{p}_{j}}),\nonumber\\
L_{\Sigma}  &  =%
{\displaystyle\sum\limits_{i=s+1}^{N}}
\mathbf{v}_{i}\mathbf{\nabla}_{i}-\sum\limits_{s+1\leq i<j\leq N}%
(\mathbf{\nabla}_{i}V_{ij})\cdot(\frac{\partial}{\partial\mathbf{p}_{i}}%
-\frac{\partial}{\partial\mathbf{p}_{j}}),\nonumber\\
\widetilde{L}_{s\Sigma}  &  =-\sum\limits_{i=1}^{s}\sum\limits_{j=s+1}%
^{N}(\mathbf{\nabla}_{i}V_{ij})\cdot(\frac{\partial}{\partial\mathbf{p}_{i}%
}-\frac{\partial}{\partial\mathbf{p}_{j}})+%
{\displaystyle\sum\limits_{i=1}^{s}}
(\mathbf{\nabla}_{i}<H_{s\Sigma}>_{\Sigma})\frac{\partial}{\partial
\mathbf{p}_{i}},\nonumber\\
L^{F}(t)  &  =\sum\limits_{i=1}^{N}L_{i}^{F}(t),L_{i}^{F}(t)=-[\mathbf{\nabla
}_{i}V_{i}(\mathbf{r}_{i},t)]\frac{\partial}{\partial\mathbf{p}_{i}%
},\nonumber\\
\mathbf{v}_{i}  &  =\mathbf{p}_{i}/m,\mathbf{\nabla}_{i}=\frac{\partial
}{\partial\mathbf{r}_{i}},V_{ij}=V_{ij}(\left\vert \mathbf{r}_{i}%
-\mathbf{r}_{j}\right\vert ). \label{31}%
\end{align}

Equation (\ref{20}) and (\ref{27}) can be rewritten and simplified if we take
into account the following operator properties%
\begin{align}
P_{s}L^{F}(t)  &  =P_{s}L^{F}(t)P_{s}=%
{\displaystyle\sum\limits_{i=1}^{s}}
L_{i}^{F}(t)P_{s},\nonumber\\
P_{s}L^{F}(t)Q_{s}  &  =0,Q_{s}L^{F}(t)P_{s}=%
{\displaystyle\sum\limits_{i=s+1}^{N}}
L_{i}^{F}(t)P_{s},\nonumber\\
P_{s}L_{s}  &  =P_{s}L_{s}P_{s}=\rho_{\Sigma}^{s}L_{s}\int...%
{\displaystyle\int}
dx^{N-s},P_{s}L_{s}Q_{s}=0,\nonumber\\
Q_{s}L_{s}P_{s}  &  =L_{s}P_{s}-\rho_{\Sigma}^{s}L_{s}\int...%
{\displaystyle\int}
dx^{N-s},\nonumber\\
P_{s}L_{\Sigma}  &  =0,P_{s}L_{\Sigma}Q_{s}=0,Q_{s}L_{\Sigma}P_{s}=L_{\Sigma
}P_{s},\nonumber\\
P_{s}\widetilde{L}_{s\Sigma}P_{s}  &  =%
{\displaystyle\sum\limits_{i=1}^{s}}
[<\mathbf{F}_{i}>_{\Sigma}^{s}-<\mathbf{F}_{i}>_{\Sigma}]\frac{\partial
}{\partial\mathbf{p}_{i}}P_{s}, \label{32}%
\end{align}
where%
\begin{equation}
\mathbf{F}_{i}=-%
{\displaystyle\sum\limits_{j=s+1}^{N}}
(\mathbf{\nabla}_{i}V_{ij}),<...>_{\Sigma}^{s}=\int...%
{\displaystyle\int}
dx^{N-s}(...\rho_{\Sigma}^{s}), \label{33}%
\end{equation}
i.e., $\mathbf{F}_{i}$ is the force acting on the $i$-particle ($i=1,...,s$)
from the the "environment" of $N-s$ particles. Here and further on, we use, as
usual, that all functions $\Phi(x_{1},\ldots,x_{N};t)$, defined on the phase
space, and their derivatives vanish at the boundaries of the configurational
space and at $\mathbf{p}_{i}=\pm\infty$.

It is evident, that initial correlations are "hidden" in the projection
operators $P_{s}$, $Q_{s}=1-P_{s}$. In the case of a small inter-particle
interaction $V_{ij}$, we can expand $P_{s}$ into series in the corresponding
small parameter. Thus, if we suppose that
\begin{equation}
<p_{i}^{2}/2m>\backsim k_{B}T>>V_{ij}, \label{34}%
\end{equation}
i.e., the interparticle interaction is small as compared to the average
particle kinetic energy. In particular, we assume, that $\beta\widetilde
{H}_{s\Sigma}$ is proportional to a small parameter. Then, approximately, in
the linear approximation in $\beta\widetilde{H}_{s\Sigma}$,
\begin{align}
P_{s}  &  =P_{s}^{1}=\frac{e^{-\beta H_{\Sigma}}(1-\beta\widetilde{H}%
_{s\Sigma})}{%
{\displaystyle\int}
...%
{\displaystyle\int}
dx^{N-s}e^{-\beta H_{\Sigma}}(1-\beta\widetilde{H}_{s\Sigma})}\int...%
{\displaystyle\int}
dx^{N-s}\nonumber\\
&  =(1-\beta\widetilde{H}_{s\Sigma})\rho_{\Sigma}\int...%
{\displaystyle\int}
dx^{N-s}\nonumber\\
&  =P-\beta\widetilde{H}_{s\Sigma}P,\nonumber\\
Q_{s}  &  =Q_{s}^{1}=Q+\beta\widetilde{H}_{s\Sigma}P, \label{35}%
\end{align}
where $P$ is the conventional projection operator (\ref{14}). It is not
difficult to verify that the operator $P_{s}^{1}$ (\ref{35}) is the projector
also. The relevant function (\ref{17}) is now given by%

\begin{align}
f_{r}^{s}(t)  &  =(1-\beta\widetilde{H}_{s\Sigma})\rho_{\Sigma}\frac{1}{V^{s}%
}F_{s}(t)\nonumber\\
&  =(1-\beta\widetilde{H}_{s\Sigma})\rho_{r}^{s}(t), \label{36}%
\end{align}
where $\rho_{r}^{s}(t)$ is defined by (\ref{15}). In this approximation, Eqs.
(\ref{32}) look like%

\begin{align}
P_{s}^{1}L^{F}(t)  &  =P_{s}^{1}L^{F}(t)P_{s}^{1}=(1-\beta\widetilde
{H}_{s\Sigma})%
{\displaystyle\sum\limits_{i=1}^{s}}
L_{i}^{F}(t)P,\nonumber\\
P_{s}^{1}L^{F}(t)Q_{s}^{1}  &  =0,Q_{s}^{1}L^{F}(t)P_{s}^{1}=(1-\beta
\widetilde{H}_{s\Sigma})\sum\limits_{i=s+1}^{N}L_{i}^{F}(t)P,\nonumber\\
P_{s}^{1}L_{s}  &  =P_{s}^{1}L_{s}P_{s}^{1}=(1-\beta\widetilde{H}_{s\Sigma
})L_{s}P,P_{s}^{1}L_{s}Q_{s}^{1}=0,Q_{s}^{1}L_{s}P_{s}^{1}=\beta
\lbrack\widetilde{H}_{s\Sigma},L_{s}]P,\nonumber\\
P_{s}^{1}L_{\Sigma}  &  =0,P_{s}^{1}L_{\Sigma}Q_{s}^{1}=0,Q_{s}^{1}L_{\Sigma
}P_{s}^{1}=L_{\Sigma}P_{s}^{1}=-\beta L_{\Sigma}\widetilde{H}_{s\Sigma
}P,\nonumber\\
P_{s}^{1}\widetilde{L}_{s\Sigma}P_{s}^{1}  &  =-(1-\beta\widetilde{H}%
_{s\Sigma})%
{\displaystyle\sum\limits_{i=1}^{s}}
<\mathbf{F}_{i}\beta\widetilde{H}_{s\Sigma}>_{\Sigma}\frac{\partial}%
{\partial\mathbf{p}_{i}}P. \label{37}%
\end{align}

Now, we are going to simplify Eq. (\ref{20}) using the relations (\ref{37})
and restricting ourselves up to the second order approximation in the
interparticle interaction. The left-hand-side of Eq. (\ref{20}) for projection
operator $P_{s}^{1}$ (\ref{35}) turns to%
\begin{equation}
\frac{\partial}{\partial t}f_{r}^{s}(t)=(1-\beta\widetilde{H}_{s\Sigma}%
)\frac{\partial}{\partial t}\rho_{r}^{s}(t). \label{38}%
\end{equation}
For $P_{s}=P_{s}^{1}$, the first term on the right-hand-side of (\ref{20}) is
as following%
\begin{align}
-P_{s}^{1}L(t)f_{r}^{s}(t)  &  =-(1-\beta\widetilde{H}_{s\Sigma}%
)[\sum\limits_{i=1}^{s}L_{i}^{F}(t)\nonumber\\
+L_{s}-\sum\limits_{i=1}^{s}  &  <\mathbf{F}_{i}\beta\widetilde{H}_{s\Sigma
}>_{\Sigma}\frac{\partial}{\partial\mathbf{p}_{i}}]\rho_{r}^{s}(t). \label{39}%
\end{align}
Taking into account that the factor $(1-\beta\widetilde{H}_{s\Sigma})$ can be
cancelled as the mutual one for (\ref{38}), (\ref{39}) and for the collision
(the second on the r.h.s.) term in (\ref{20}), we can see that the first two
terms on the r.h.s. of (\ref{39}) are the conventional terms which result from
the term $-PL(t)\rho_{r}^{s}(t)$ of Eq. (\ref{11}), and the third term on the
r.h.s. of (\ref{39}) represents the correction due to initial correlations
(i.e.,to the deviation of the projector $P_{s}^{1}$ from the conventional one
$P$).

Employing Eqs. (\ref{37}), the second on the r.h.s. of (\ref{20}) collision
term can be rewritten as
\begin{align}
C(t)  &  =\int\limits_{0}^{t}P_{s}L(t)U_{Q_{s}}(t,\tau)Q_{s}L(\tau)f_{r}%
^{s}(\tau)d\tau=\int\limits_{0}^{t}P_{s}^{1}\widetilde{L}_{s\Sigma}%
U_{Q_{s}^{1}}(t,\tau)Q_{s}^{1}[\widetilde{L}_{s\Sigma}+L^{F}(\tau)\nonumber\\
&  +L_{s}+L_{\Sigma}]P_{s}^{1}f_{r}^{s}(\tau)d\tau. \label{40}%
\end{align}
As we will see later, the third and fourth terms in the r.h.s. of (\ref{40})
appear due to initial correlations, while the first and the second ones are
the same as for Eq. (\ref{11}) with $\rho_{i}(0)=0$.

For simplicity, from now on, we will consider the case of a weak external
field (linear response regime), i.e., only the terms of the order
$(V_{ij})^{m}V_{i}$ $(m=0,1)$ will be accounted for. Then, in the second order
approximation for the collision integral in the weak interaction and for a
weak external field, $U_{Q_{s}^{1}}(t,\tau)$ could be taken in the zero order
approximation in $V_{ij}$
\begin{align}
U_{Q_{s}^{1}}^{0}(t,\tau)  &  =\exp[-(L_{s}^{0}+L_{\Sigma}^{0})(t-\tau
)],\nonumber\\
L_{s}^{0}  &  =%
{\displaystyle\sum\limits_{i=1}^{s}}
\mathbf{v}_{i}\mathbf{\nabla}_{i},L_{\Sigma}^{0}=%
{\displaystyle\sum\limits_{i=s+1}^{N}}
\mathbf{v}_{i}\mathbf{\nabla}_{i} \label{41}%
\end{align}
taking into account that the other factors under the integral in (\ref{40})
are already of the second order in the interaction. Then, for any function of
the particles coordinates $\Phi(\mathbf{r}_{1},..,\mathbf{r}_{N})$
\begin{align}
&  \exp[-(L_{s}^{0}+L_{\Sigma}^{0})\tau\mathbf{]}\Phi(\mathbf{r}%
_{1},..,\mathbf{r}_{N})\nonumber\\
&  =\Phi(\mathbf{r}_{1}-\mathbf{v}_{1}\tau,...,\mathbf{r}_{N}-\mathbf{v}%
_{N}\tau). \label{42}%
\end{align}
To remain within the adopted accuracy, the distribution $\rho_{\Sigma}$ in
(\ref{39}) and (\ref{40}) should be taken in the zero approximation in the
interaction%
\begin{equation}
\rho_{\Sigma}^{0}=\frac{\exp(-\beta%
{\displaystyle\sum\limits_{i=s+1}^{N}}
\frac{\mathbf{p}_{i}^{2}}{2m})}{\prod\limits_{i=s+1}^{N}\int d\mathbf{p}%
_{i}d\mathbf{r}_{i}\exp(-\beta%
{\displaystyle\sum\limits_{i=s+1}^{N}}
\frac{\mathbf{p}_{i}^{2}}{2m})}=\frac{\prod\limits_{i=s+1}^{N}\exp(-\beta
\frac{\mathbf{p}_{i}^{2}}{2m})}{V^{N-s}\prod\limits_{i=s+1}^{N}[\int
d\mathbf{p}_{i}\exp(-\beta\frac{\mathbf{p}_{i}^{2}}{2m})]}. \label{43}%
\end{equation}

The non-Markovian Eq. (\ref{20}) can be turned to a time-local form if we take
$f_{r}^{s}(\tau)$ in (\ref{40}) in the zero order approximation
\begin{equation}
f_{r}^{s}(\tau)=e^{L_{s}^{0}(t-\tau)}f_{r}^{s}(t) \label{44}%
\end{equation}
in order to remain in the adopted second order in interaction approximation.
Now, after changing the variable $t-\tau\rightarrow\tau$ under the integral,
the approximate collision term is%
\begin{equation}
C(t)=\int\limits_{0}^{t}P_{s}^{1}\widetilde{L}_{s\Sigma}e^{-(L_{s}%
^{0}+L_{\Sigma}^{0})\tau}Q_{s}^{1}[\widetilde{L}_{s\Sigma}+L^{F}(t-\tau
)+L_{s}+L_{\Sigma}]P_{s}^{1}e^{L_{s}^{0}\tau}f_{r}^{s}(t)d\tau. \label{45}%
\end{equation}

In the same way we can simplify Eq. (\ref{28}) in the case of a small
interparticle interaction (see (\ref{34})) and a small external field by using
the approximate value $P_{s}^{1}$ (\ref{35}) for the the projection operator
$P_{s}$ and restricting ourselves to the second order in the interaction and
to the terms of the order $(V_{ij})^{m}V_{i}$ $(m=0,1)$. Using (\ref{37}), we
see that the l.h.s. and the first term on the r.h.s. of (\ref{28}) are given
by (\ref{38}) and (\ref{39}). From relations (\ref{37}), (\ref{41}) and
(\ref{43}), in the adopted approximation, the collision term (the second on
the r.h.s. of (\ref{28})) reads as%
\begin{equation}
C(t)=\int\limits_{0}^{t}d\tau P_{s}^{1}\widetilde{L}_{s\Sigma}e^{-(L_{s}%
^{0}+L_{\Sigma}^{0})\tau}Q_{s}^{1}[\widetilde{L}_{s\Sigma}+L^{F}(t-\tau
)+L_{s}+L_{\Sigma}]P_{s}^{1}e^{(L_{s}^{0}+L_{\Sigma}^{0})\tau}f_{r}^{s}(t).
\label{46}%
\end{equation}
where $f_{r}^{s}(t)$ should be approximated as $f_{r}^{s}(t)=\rho_{\Sigma}%
^{0}\frac{1}{V^{s}}F_{s}(t)$. Now one can see that $e^{L_{\Sigma}^{0}\tau
}f_{r}^{s}(t)=f_{r}^{s}(t)$, and (\ref{46}) coincides with (\ref{45}).

\section{Equation for a one-particle distribution function}

A one-particle distribution function $F_{1}(x_{1},t)$ is often mainly of
interest. This case is less involved, although the developed formalism can
also be specified for two-, three-, and more-particle distribution functions.
Let us write down the terms of Eq. (\ref{28}), given by (\ref{38}), (\ref{39})
and (\ref{45}) for $s=1$. Omitting the common factor $(1-\beta\widetilde
{H}_{1\Sigma})\rho_{\Sigma}/V$ and using Eqs. (\ref{29}) - (\ref{31}),
(\ref{35}), and (\ref{41}) - (\ref{43}), we obtain%
\begin{align}
L_{1}^{F}(t)\rho_{r}^{1}(x_{1},t)  &  =-[\mathbf{\nabla}_{1}V_{1}%
(\mathbf{r}_{1},t)]\frac{\partial}{\partial\mathbf{p}_{1}}F_{1}(x_{1}%
,t),\nonumber\\
L_{1}\rho_{r}^{1}(x_{1},t)  &  =[\mathbf{v}_{1}\mathbf{\nabla}_{1}%
-(\mathbf{\nabla}_{1}<H_{1\Sigma}>_{\Sigma})\frac{\partial}{\partial
\mathbf{p}_{1}}]F_{1}(x_{1},t),\nonumber\\
&  <\mathbf{F}_{1}\beta\widetilde{H}_{1\Sigma}>_{\Sigma}\frac{\partial
}{\partial\mathbf{p}_{1}}\rho_{r}^{1}(x_{1},t)=-\beta n\int dr_{2}%
(\mathbf{\nabla}_{1}V_{12})V_{12}\frac{\partial}{\partial\mathbf{p}_{1}}%
F_{1}(x_{1},t), \label{47}%
\end{align}
where $n=N/V$ (here, as usual, the limiting procedure, $N\rightarrow\infty$,
$V\rightarrow\infty$ with $n$ remaining fixed is assumed). Here we also used
that in the first approximation in $V_{ij}$
\begin{align}
&  <H_{1\Sigma}>_{\Sigma}=\prod\limits_{i=2}^{N}\int d\mathbf{r}_{i}\int
d\mathbf{p}_{i}\rho_{\Sigma}^{0}\sum\limits_{j=2}^{N}V_{1j}\nonumber\\
&  =n\int d\mathbf{r}_{2}V_{12}. \label{48}%
\end{align}
\qquad

The first term in the collision integral (\ref{46}) for the case under
consideration ($s=1$) is as following
\begin{align}
C_{L}(x_{1},t)  &  =\int\limits_{0}^{t}d\tau P_{s}^{1}\widetilde{L}_{s\Sigma
}U_{Q_{s}^{1}}(\tau)Q_{s}^{1}\widetilde{L}_{s\Sigma}P_{s}^{1}e^{L_{s}^{0}\tau
}f_{r}^{1}(t)\nonumber\\
&  =n\int\limits_{0}^{t}d\tau%
{\displaystyle\int}
d\mathbf{p}_{2}\mathbf{\partial}_{1}G_{L}(x_{1},\mathbf{g}_{12}\mathbf{;}%
\tau)(\partial_{12}+\frac{\tau}{m}\mathbf{\nabla}_{1})\rho_{\Sigma}^{(2)}%
F_{1}(x_{1},t), \label{49}%
\end{align}
where%
\begin{align}
G_{L}(x_{1},\mathbf{g}_{12}\mathbf{;}\tau)  &  =%
{\displaystyle\int}
d\mathbf{r}_{2}\mathbf{F}_{12}(0)\mathbf{F}_{12}(\tau),\nonumber\\
\mathbf{F}_{12}(\tau)  &  =-\mathbf{\nabla}_{1}V(\mathbf{r}_{1}-\mathbf{r}%
_{2}-\mathbf{g}_{12}\tau),\mathbf{g}_{12}\mathbf{=v}_{1}-\mathbf{v}%
_{2},\nonumber\\
\rho_{\Sigma}^{(2)}  &  =\frac{\exp(-\beta\frac{p_{2}^{2}}{2m})}{\int
d\mathbf{p}_{2}\exp(-\beta\frac{p_{2}^{2}}{2m})},\partial_{1}=\frac{\partial
}{\partial\mathbf{p}_{1}},\partial_{12}=\frac{\partial}{\partial\mathbf{p}%
_{1}}-\frac{\partial}{\partial\mathbf{p}_{2}}. \label{50}%
\end{align}

The second term in (\ref{46}), related to the external field which acts on the
particles of a reservoir and due to the interparticle interaction influences
the targeted particle, looks as%
\begin{align}
C_{F}(x_{1},t)  &  =n\int\limits_{0}^{t}d\tau%
{\displaystyle\int}
d\mathbf{p}_{2}G_{F}(\mathbf{r}_{1},\mathbf{v}_{2};\tau)\partial_{2}%
\rho_{\Sigma}^{(2)}\partial_{1}F_{1}(x_{1},t),\nonumber\\
G_{F}(\mathbf{r}_{1},\mathbf{v}_{2};\tau)  &  =%
{\displaystyle\int}
d\mathbf{r}_{2}\mathbf{F}_{12}(0)\mathbf{F}_{2}^{F}(\tau),\nonumber\\
\mathbf{F}_{2}^{F}(\tau)  &  =-\mathbf{\nabla}_{2}V_{2}(\mathbf{r}%
_{2}-\mathbf{v}_{2}\tau,t-\tau),\partial_{2}=\frac{\partial}{\partial
\mathbf{p}_{2}}. \label{51}%
\end{align}
It is not difficult to see, that in the case of a homogeneous external force
(e.g.,$V_{i}(\mathbf{r}_{i},t)=-e\mathbf{E(}t\mathbf{)r}_{i}$), $\mathbf{F}%
_{2}^{F}(\tau)$ does not depend on $\mathbf{v}_{2}$ and, therefore,
$C_{F}(x_{1},t)=0$ (see definition for $\rho_{\Sigma}^{(2)}$ (\ref{50})).

The third term in (\ref{46}), related to $L_{s}$ ($s=1$), looks like%
\begin{align}
C_{1}(x_{1},t)  &  =n\beta\int d\mathbf{p}_{2}\partial_{1}\mathbf{v}_{1}%
G_{1}(x_{1},\mathbf{g}_{12};t)\rho_{\Sigma}^{(2)}F_{1}(x_{1},t),\nonumber\\
G_{1}(x_{1},\mathbf{g}_{12};t)  &  =-\int\limits_{0}^{t}d\tau%
{\displaystyle\int}
d\mathbf{r}_{2}\mathbf{F}_{12}(0)[\mathbf{\nabla}_{1},V(\mathbf{r}%
_{1}-\mathbf{r}_{2}-\mathbf{g}_{12}\tau)]. \label{52}%
\end{align}
In the space-homogeneous case, when $F_{1}(x_{1},t)=\varphi(\mathbf{p}_{1}%
,t)$, $G_{1}(x_{1},\mathbf{g}_{12};t)=\int\limits_{0}^{t}d\tau G_{L}%
(x_{1},\mathbf{g}_{12}\mathbf{;}\tau)$ (the second term of the commutator in
(\ref{43}) $V(\mathbf{r}_{1}-\mathbf{r}_{2}-\mathbf{g}_{12}\tau)\mathbf{\nabla
}_{1}\varphi(\mathbf{p}_{1},t)=0$).

And the last term in the collision integral (\ref{46}), stipulated by
$L_{\Sigma}$, at $s=1$ is%
\begin{equation}
C_{2}(x_{1},t)=-n\beta\int\limits_{0}^{t}d\tau\int d\mathbf{p}_{2}\partial
_{1}\mathbf{v}_{2}G_{L}(x_{1},\mathbf{g}_{12};\tau)\rho_{\Sigma}^{(2)}%
F_{1}(x_{1},t), \label{53}%
\end{equation}
where we used that $\mathbf{\nabla}_{2}V_{12}=-\mathbf{\nabla}_{1}V_{12}$.

Collecting all above obtained results, we finally obtain from (\ref{46}) the
following equation for a one-particle distribution function in the second
order approximation in the inter-particle interaction%
\begin{align}
\frac{\partial}{\partial t}F_{1}(x_{1},t)  &  =\{[\mathbf{\nabla}_{1}%
U_{1}(\mathbf{r}_{1},t)]+n(\mathbf{\nabla}_{1}\int d\mathbf{r}_{2}%
V_{12})-\beta n\int dr_{2}(\mathbf{\nabla}_{1}V_{12})V_{12}\}\frac{\partial
}{\partial\mathbf{p}_{1}}F_{1}(x_{1},t)\nonumber\\
&  -\mathbf{v}_{1}\mathbf{\nabla}_{1}F_{1}(x_{1},t)+[C_{L}(x_{1}%
,t)+C_{F}(x_{1},t)+C_{1}(x_{1},t)+C_{2}(x_{1},t)] \label{54}%
\end{align}
Equation (\ref{54}) is the main result of this section. We stress again that
it is a linear equation strictly obtained in the second order approximation in
$V_{ij}$ and is valid for all timescales. The collision integral $C_{L}%
(x_{1},t)$ coincides with the corresponding collision integral in the
nonlinear equation for inhomogeneous system of classical particles (see
\cite{Balescu}) if in the latter, the distribution function for the particle,
with which the targeted particle collides, is replaced by the equilibrium
distribution function for this particle $\rho_{\Sigma}^{(2)}$. However, for
such a coincidence, the integral over $d\tau$ should be extended to infinity.
It can be done, if the interaction is rather a short-range one and if for a
timescale%
\begin{equation}
t>t_{cor}\sim r_{cor}/v \label{55}%
\end{equation}
the force acting on the particle vanishes ($\mathbf{F}_{12}(t)=0$), where
$r_{cor}$ is a radius of the inter-particle interaction $V_{ij}$ and $v$ is
the average particle velocity. The third term in the first line of (\ref{54})
(proportional to $\beta$), as well as the collision integrals $C_{1}(x_{1},t)$
and $C_{2}(x_{1},t)$, appear due to the initial correlation contribution.

Equation (\ref{54}) acquires more simple form in the space-homogeneous case
when an external force is a coordinate-independent one and $F_{1}%
(x_{1},t)=\varphi(\mathbf{p}_{1},t)$. In this case Eq. (\ref{54}) reads as follows%

\begin{align}
\frac{\partial}{\partial t}\varphi(\mathbf{p}_{1},t)  &  =[\mathbf{\nabla}%
_{1}U_{1}(\mathbf{r}_{1},t)\frac{\partial}{\partial\mathbf{p}_{1}}+n%
{\displaystyle\int}
d\mathbf{p}_{2}\mathbf{\partial}_{1}\overline{G}_{L}(\mathbf{g}_{12}%
\mathbf{,}t)\partial_{12}\rho_{\Sigma}^{(2)}]\varphi(\mathbf{p}_{1}%
,t)\nonumber\\
&  +n\beta%
{\displaystyle\int}
d\mathbf{p}_{2}\mathbf{\partial}_{1}\mathbf{g}_{12}\overline{G}_{L}%
(\mathbf{g}_{12}\mathbf{,}t)\rho_{\Sigma}^{(2)}\varphi(\mathbf{p}%
_{1},t),\nonumber\\
\overline{G}_{L}(\mathbf{g}_{12}\mathbf{,}t)  &  =\int\limits_{0}^{t}d\tau%
{\displaystyle\int}
d\mathbf{r[\nabla}_{1}V(\mathbf{r})][\mathbf{\nabla}_{1}V(\mathbf{r-g}%
_{12}\tau)]. \label{56}%
\end{align}

Note, that Eqs. (\ref{54}) and (\ref{56}) are the reversible in time ones.
They become irreversible in the scale limit $t\rightarrow\infty$ (when the
latter exists), e.g., in the case of the rapid vanishing of the interparticle interaction.

It is interesting to compare Eq. (\ref{56}) with the Landau irreversible in
time nonlinear equation for a homogeneous gas of classical particles, which
can be derived from the the Boltzmann equation for the weak inter-particle
interaction (see, e.g. \cite{Balescu}). The collision integral in the linear
Eq. (\ref{56}) will coincide with the Landau collision integral only if we
disregard the contribution of initial correlations (the third term on the
r.h.s. of Eq. (\ref{56})) and in the Landau equation replace the distribution
function for the second targeted particle $\varphi(\mathbf{p}_{2},t)$ by the
equilibrium distribution function $\rho_{\Sigma}^{(2)}$. However, this
coincidence will be valid only on the timescale (\ref{55}), when the integral
in $\overline{G}_{L}(\mathbf{g}_{12}\mathbf{,}t)$ can be extended to infinity.
At the same time, the linear collision integral of Eq. (\ref{56}) at
$t\rightarrow\infty$ can be rewritten in the form coincided with that of the
Fokker-Planck equation (see \cite{Balescu}). We would like to stress again,
that linear Eqs. (\ref{54}) and (\ref{56}) are obtained strictly in the second
order of the perturbation theory from the homogeneous exact GMEs (\ref{20})
and (\ref{27}), and that they are valid on the arbitrary timescale.

\section{Conclusion}

For a system of $N$ ($N>>1$) classical particles driven from the initial
equilibrium state by an external field, the exact new time-convolution and
time-convolutionless homogeneous (completely closed) linear GMEs for an
$s$-particle ($s<N$) distribution functions valid on the arbitrary timescale
have been obtained. It has become possible due to the special projection
operator (\ref{16}) which accounts for initial correlations in the initial
equilibrium distribution function of the whole system. No assumption like
"molecular chaos" (factorized initial state) or Bogoliubov's principle of
weakening of initial correlations, which results in converting the linear
Liouville equation into the nonlinear evolution equation, has been used. By
expanding the kernels of the obtained equations in the perturbation series in
the interparticle interaction, the corresponding equation in the second order
approximation for a weak interaction and small external field is obtained.
This equation is specified for a one-particle distribution function in the
space-inhomogeneous (Eq. (\ref{54})) and space-homogeneous (Eq. (\ref{56}))
cases. Both these equations contain contribution of initial correlations which
have not been conveniently disregarded. The collision integral in (\ref{56})
is the linear Landau (Fokker-Planck) collision integral on the large timescale.

\bigskip

\end{document}